\documentstyle[psfig]{l-aa}
\def\etal{{\it et al.\,}}

\def\ie{{\it i.e.\/}}

\title{ The Optical Transient of GRB970228, 16 hours after the burst }

\author{F.~Pedichini\inst{1}
\and A.~Di~Paola\inst{1}
\and L.~Stella\inst{1,}\inst{2}
\and R.~Buonanno\inst{1}
\and A.~Boattini\inst{1}
\and G.~Gandolfi\inst{3}
\and E.~Costa\inst{4}
\and M.~Feroci\inst{4}
\and L.~Piro\inst{4}
\and D.~Dal Fiume\inst{5}
\and F.~Frontera\inst{5,}\inst{6}
\and L.~Nicastro\inst{5}
\and E.~Palazzi\inst{5}
\and J.~Heise\inst{7}
\and J.~in't~Zand\inst{7} 
\and M.~Vietri\inst{8} }

\thesaurus{13 (13.07.1; 13.07.2)}

\begin{document}
\institute{
{Osservatorio Astronomico di Roma, Via dell'Osservatorio 2, I-00040 
Monteporzio Catone (Roma), Italy; \\ e-mail: pedik, dipaola,  
stella, buonanno, boattini@coma.mporzio.astro.it}
\and 
{Affiliated to the International Center for Relativistic Astrophysics}
\and 
{Beppo-SAX Scientific Data Centre, Via Corcolle 19, I-00131 Roma, Italy; 
\\ e-mail: gandolfi@napa.sdc.asi.it }
\and 
{Istituto Astrofisica Spaziale, C. N. R., Via E. Fermi 21, I-00044
Frascati (Roma), Italy; \\ e-mail: 
costa, feroci, piro@saturn.ias.fra.cnr.it}
\and 
{Istituto TeSRE, C. N. R., Via P. Gobetti 101, I-40129 Bologna , Italy; 
\\ e-mail: dalfiume, filippo, nicastro, palazzi@tesre.bo.cnr.it}
\and 
{Dipartimento di  Fisica, Universit\`a di Ferrara, Via Paradiso 12, 
I-44100 Ferrara, Italy;}
\and 
{Space Research Organisation in the Netherlands, Sorbonnelaan 2, NL-3584 CA 
Utrecht, The Netherlands; \\ e-mail: jheise, jeanz@sron.ruu.nl}
\and 
{Dipartimento di Fisica, Universit\`a di Roma 3, 
Via della Vasca Navale 84, I-00146 Roma, Italy; \\ e-mail: 
 vietri@corelli.fis.uniroma3.it}
}

\date{Received ; Accepted}

\maketitle
\label{sampout}
                         
\begin{abstract}
Until recently the positions of gamma ray bursts
were not sufficiently well known within a short timescale
to localize and identify them with known celestial sources.
Following the historical detection of the X--ray afterglow of the 
burst GRB970228, extending from $30$~s to $3$~ days after the main peak, 
by the Beppo-SAX satellite (Costa \etal \cite{costa}), and that of an optical 
transient $\sim 21$~hr after the burst (van Paradijs \etal 1997, hereafter 
\cite{vp}), we report here the detection of the same optical transient, 
in images obtained $\sim 4.5$~hr hours 
earlier than in \cite{vp}.\\
\keywords{ X-ray -- Gamma Ray Burst -- Optical Transient }
\end{abstract}

Multiwavelength detection of GRB 
counterparts has been made possible recently by Beppo-SAX, the 
Italian-Dutch satellite for X-ray astronomy (Piro \etal \cite{piro}): several gamma
ray bursts (GRBs) have been localized with an accuracy of $\sim  3$~arcmin
within a few hours from their occurrence. Well-targeted follow up 
observations of GRBs from X-ray to radio wavelengths have thus been made 
within an unprecedented time lapse of less than a day (Paczy\'nski \& Wijers
\cite{pac}). In the case of 
GRB970228, X--ray observations with the Beppo-SAX Narrow Field Instruments
have allowed further refinement of the position to within $45$~arcsec, 
and optical searches have found a faint but 
undisputable Optical Transient within this small error box.

The detection of the X-ray afterglow is 
interpreted as evidence for highly relativistic expansion of matter 
ejected from an as yet undetermined source (Vietri \cite{vietria}), within the 
framework of the fireball model (M\`esz\`aros \etal \cite{mesz}).
The detection and monitoring of the optical counterpart is of 
paramount importance, because it allows source localization with highest 
precision and searches for a possible host. Depending on peak flux levels,
spectroscopic techniques can be employed to harness much needed information
on the distance scale, a still elusive topic. Also, a determination of the
time--dependence of the counterpart fading can provide tests of different 
GRB models.

In response to the alert from the Beppo-SAX Team, we observed the field  
containing the (then) $\sim 3$~arcmin accurate position of GRB970228 
with the Rome Astronomical Observatory's $0.60$~m Schmidt telescope at 
Campo Imperatore on Feb $28.8$, Mar $4.8$ and Mar $12.8$ UT. The 
telescope was equipped with a $2k\times 2k$ CCD camera covering about 
$1$~deg$^2$  at a scale of $1.67$~arcsec/pixel. No filter was used in 
front of the detector in order to achieve maximum sensitivity within the 
relatively short exposure allowed by the field location, just beyond the
local meridian  at sunset. The resulting band peaks at $\sim 700$~nm with 
a FWHM of $\sim 300$~nm, corresponding roughly to $0.5F_V+F_R+F_I$ 
when referred 
to the Bessel filter set. Each image was composed by dithering three frames
taken in sequence after a small shift of the telescope pointing direction; 
each frame was exposed for $\sim 500$~s. The FWHM of the point spread function 
were $\sim 4,\; 4 $~and $3$~arcsec on Feb $28$, Mar $4$ and Mar $12$, 
respectively. The first observation took place between Feb $28.795$ and 
$28.827$ UT, starting only $16.1$~hr after the onset of GRB970228 (Feb 
$28.124$ UT $\equiv t_\circ$ (Costa \etal \cite{costa})), \ie $\sim 4.5$~hours 
earlier than the  observation that led to the discovery of the Optical 
Transient in the Beppo-SAX error box of GRB970228 (\cite{vp}).

Our image (fig.~1a) revealed  an object at a position of RA 
$05h\ 01m\ 46.63\pm 0.04s$ and DEC $11^\circ \ 46'\ 54.7\pm 1.0"$, 
consistent with the position of the Optical Transient (OT) reported in 
\cite{vp} (we give maximum uncertainties throughout this 
correspondence). Some $3$~arcsec to the SW at RA $05h\ 01m\ 46.52\pm 0.04s$
and DEC $11^\circ \ 46'\ 53.0\pm 1.0"$, the nearby KM star reported in 
\cite{vp}, Groot \etal (\cite{groot}), Tonry \etal (\cite{tonry}) was visible but partially 
blended with the OT. Then, to extract the net fluxes of both sources, we 
applied an iterative debiasing routine showing that the transient was 
$1.6\pm0.5$~mag brighter than the KM star in our band. Aperture photometry 
on four nearby field stars, supposed constant and labelled $1,2,3,4$ in 
the figure, set the zero point of the magnitudes. This reference set 
showed a differential maximum error between different nights of less than 
0.1 mag.        

\begin{figure}
\centerline{
\psfig{figure=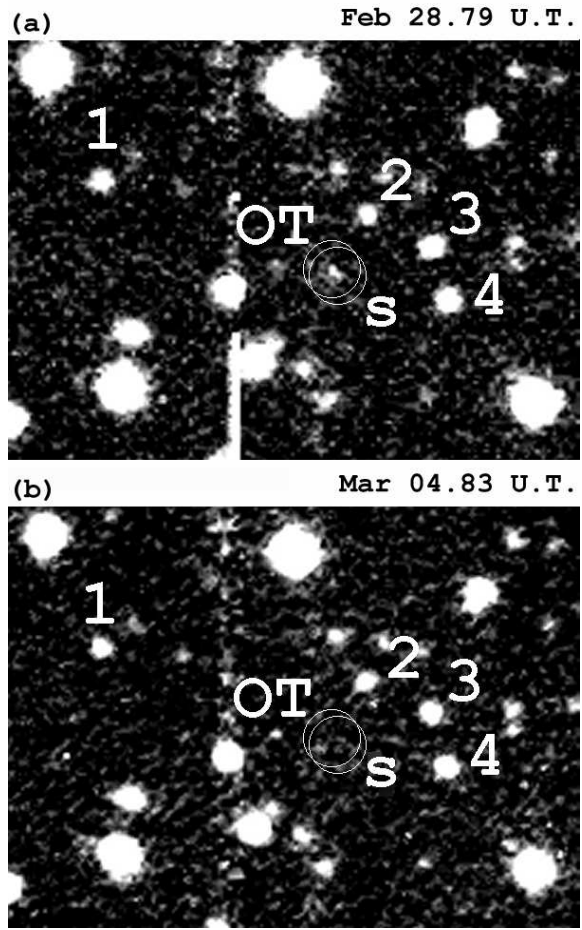,width=7.6cm,height=12.5cm}
}
\caption[]{CCD images of the field of GRB970228 taken at the 
0.60~m Schmidt telescope at Campo Imperatore. The numbers indicate 
the reference stars, used to set the photometry zero point. The centers
of the circles indicate the location of the Optical Transient (labelled
OT) and of the KM star (labelled S). Each image covers a field of $\approx 
2.8\times1.9$~arcmin$^2$. North and East are top and left, respectively.}
\end{figure}

Further analysis on a short time scale (using separately the three frames 
of Feb $28$) did not identify variability of the Optical Transient during 
the $50$ minutes observation run, with a 3$\sigma$ upper limit of about $0.7
$~mag. During our second observation (Mar $4.832$ - $4.864$ UT, see 
fig.~1b) the Optical Transient was not detected, while the nearby KM 
star was still observed at a level compatible with that of the first 
observation ($0.5$ mag uncertainty). An upper limit to the flux of the 
Optical Transient was derived and calibrated relative to our zero point 
using a detection limit equal to the $3\sigma$ sky background noise. We 
found that the Optical Transient decreased by at least $2.7$~mag in our 
band between Feb $28.8$ and Mar $4.8$. This is consistent with a power 
law fading of the flux $\propto (t-t_\circ)^{-\alpha}$ with $\alpha\geq 1.3$ .
The images we obtained on Mar $12.8$ had a substantially higher sky 
background due to the presence of the rising moon near the field of view 
and their limiting sensitivity was insufficient to detect either the nearby KM 
star or the Optical Transient. In order to convert our measurements 
to standard flux units we adopted the following procedure: we approximated our 
band with the following linear combination of the standard Bessel filters,
$0.5F_V + F_R + F_I$. To transform our signal to $\mu Jy$ we referred 
the photometry to the magnitudes reported in \cite{vp} for the KM 
dwarf, and dereddened it by means of the mean galactic absorption in the 
direction of GRB970228, $A_V\simeq 0.4$,  $A_R\simeq 0.3$,  $A_I\simeq 
0.2$. For the star, $V-I\simeq 2.4$, implying 
a M2 spectral type and, in turn, $V-R \approx 1.2$. 
Assuming for the 
Optical Transient a power law spectrum with an energy index of 0.4  
(consistent with the V and I measurements in \cite{vp})
we derive
a dereddened flux at $700\; nm$ of $50\pm 25 \mu Jy$ from our Feb.~28.8 image. 
By comparison, the I and V measurements obtained $\sim 4.5$~hours later 
(\cite{vp}) correspond to an Optical Transient flux 
of $\approx 18 \mu Jy$ at $700\; nm$. 
If the flux decreases with time according to
the law $\propto (t-t_\circ)^{-1.3}$, on the basis of the flux measured by
\cite{vp} we expect a flux of $\approx 27\; \mu Jy$ at the
time of our observations: this compares favourably with our measure of
$50\pm 25 \; \mu Jy$.

In short, we have confirmed the detection of the Optical Transient at a 
flux level comparable to that reported in \cite{vp}, extended 
backward by $\sim 4.5$ hours the time range over which the transient was 
detected (see also Guarnieri \etal \cite{guarn}), 
shown that the time law for its fading is consistent with that of 
Galama \etal (\cite{galama})
and proved the value of 
small easily accessible telescopes in the business at hand. 

If the time--delay between the beginning of the X--ray afterglow 
and the onset of the optical is sufficiently short, 
the Optical Transient following GRBs may reach $m_V \approx 17$ a few 
hours after the burst, a time scale 
long enough for the detection of the burst in the Wide Field Cameras of 
Beppo-SAX to be analyzed and relayed to optical observatories. While these 
flux levels are unlikely to be accessible to all bursts, as shown by the 
cases of GRB970111 and GRB970402, detection of even a handful of them 
would allow spectroscopic investigations establishing their 
distance scale by, {\it e.g.}, showing absorption lines of Galactic or 
Extragalactic origin, like in GRB970508 (Metzger \etal \cite{metzger}),
or traces of the Lyman--$\alpha$ forest.

Further testing of the fireball model is also made possible by optical 
observations: in fact, the time-delay between the X--ray and optical 
onsets is predicted in this model. This is because, at least initially,
most non--thermal electrons will be emitting at shorter wavelengths
than the optical ones, so that the onset of optical emission must 
wait for the peak of synchrotron emission to enter the optical
waveband. This is especially sensitive to the time--evolution
of the bulk Lorenz factor, which might be considerably affected
by the presence (or lack thereof) of surrounding matter. Together
with the time--evolution of the X--ray afterglow luminosity, 
this may yield valuable information on the environments in which
GRBs go off (Vietri \cite{vietrib}).

\end{document}